\documentclass[sigconf]{acmart}
\usepackage{enumitem}
\AtBeginDocument{%
  }

\copyrightyear{2026}
\acmYear{2026}
\setcopyright{cc}
\acmConference[HCOMP 2026]{2026 ACM Conference on Human-AI Complementarity and Alignment}{September 27--30, 2026}{Alexandria, VA, USA}
\acmBooktitle{2026 ACM Conference on Human-AI Complementarity and Alignment (HCOMP 2026), September 27--30, 2026, Alexandria, VA, USA}

\acmDOI{}
\acmISBN{}

\begin{document}

\title{SPEAR: Five Principles for Interactive Human-Agent Alignment}

\settopmatter{printacmref=false}
\author{Tao Long}
\email{long@cs.columbia.edu}
\affiliation{%
  \institution{Columbia University}
  \city{New York City}
  \state{New York}
  \country{USA}
}

\author{Lydia B. Chilton}
\email{chilton@cs.columbia.edu}
\affiliation{%
  \institution{Columbia University}
  \city{New York City}
  \state{New York}
  \country{USA}
}
\renewcommand{\shortauthors}{Long and Chilton}
\newcommand{\spear}[1]{\underline{\textbf{\textit{#1}}}}

\begin{abstract}
Recent AI alignment work often frames alignment as a pre-deployment optimization problem: collect human feedback, learn preferences or principles, fine-tune the model, and deploy an aligned system. This framing has produced major progress, but it under-specifies what happens once AI systems act as agents on users' behalf in situated, long-term, and social contexts. This position paper reframes human-agent alignment as an ongoing interaction design problem. We propose \textbf{SPEAR}, five pillars of interactive alignment: \underline{S}pecification (how people express intent and establish shared understanding), \underline{P}rocess (how agents decide when to act, ask, defer, or pause), \underline{E}valuation (how people judge whether agents succeeded), \underline{A}daptation (how agents adapt to users over repeated use), and \underline{R}ecalibration (how people adapt their trust, expectations, and behavior in response to agents).
\end{abstract}





\maketitle

\section{Introduction}

AI alignment is often defined in machine learning as the problem of \textit{training models to act in accordance with human intentions}~\cite{ouyang2022training}. This framing has produced a powerful family of pre-deployment alignment methods. Reinforcement learning from human feedback uses human preference comparisons to train reward models~\cite{christiano2017deep}; instruction tuning combines human demonstrations and preference data to improve helpfulness and instruction following~\cite{ouyang2022training}; and constitutional AI encodes explicit normative principles into model behavior~\cite{bai2022constitutional}. Alignment is therefore commonly operationalized as a linear pipeline: collect human feedback, learn or encode an objective, fine-tune the model, and deploy an aligned system.

This model-centered framing is necessary, but increasingly insufficient for interactive AI systems, and especially for AI agents that plan, use tools, and act on users' behalf. In deployed use, human intentions are rarely complete, stable, or fully specifiable in advance. Users may not know exactly what they want until they see what the system produces. They may refine goals across turns, discover new uses, change preferences, or develop new expectations as they appropriate the system into everyday work. In this setting, alignment cannot be understood only as a property encoded into model weights before deployment. Instead, alignment emerges through interaction after deployment.

Recent HCI work makes this shift explicit. Terry et al.~\cite{terry2023interactive} argue that AI alignment should be understood as an ongoing, multi-stage interactive process consisting of specification alignment, process alignment, and evaluation alignment. Specification alignment concerns how humans and AI systems establish a shared understanding of the desired goal. Process alignment concerns how humans and AI systems coordinate the actions and procedures used to achieve that goal. Evaluation alignment concerns how users determine whether the AI's behavior or output actually succeeded. This framework reframes alignment from a question of output correctness alone to a question of interaction design: how should systems communicate, coordinate, and support user judgment throughout use?
Chen et al. and Shen et al. extend this view by arguing that human-AI alignment is iterative and bidirectional~\cite{chen_align_2026,10.1145/3706599.3716291}. AI systems adapt to humans, but humans also adapt to AI. Users update their mental models, calibrate or miscalibrate trust, change behaviors, form social expectations, and sometimes become over-reliant on AI systems. Thus, alignment does not only concern whether AI follows human intent. It also concerns how interaction with AI reshapes human judgment, agency, and social behavior over time.

Building on these perspectives, this position paper synthesizes an \textbf{\textit{HCI definition of interactive human-AI alignment}}:
\begin{quote}
\emph{Alignment is the ongoing, interactive process through which humans and AI systems develop, maintain, and repair shared understanding, appropriate agency, and calibrated reliance over time.
}
\end{quote}
This definition reframes alignment as an interaction-design problem. Rather than asking only how to train a model to satisfy human preferences, it asks how humans and AI systems continuously communicate, coordinate, evaluate, adapt, and co-adapt in situated use. These questions become sharper for agents, which act rather than only respond; we therefore focus on \textbf{\textit{{human-agent} alignment}} and ask five interactive alignment questions as \textbf{\underline{\textit{SPEAR}}}:

\begin{itemize}[left=0pt]
    \item \textbf{\textit{\underline{S}pecification alignment:}} How do people express intent, and how do people and agents reach a shared understanding?
    \vspace{2px}
    \item \textbf{\textit{\underline{P}rocess alignment:}} How does an agent decide when to act, ask, defer, or pause in ways that match what the user would want?
        \vspace{2px}
    \item \textbf{\textit{\underline{E}valuation alignment:}} How do people and agents agree on what counts as success?
        \vspace{2px}
    \item \textbf{\textit{\underline{A}daptation alignment:}} What should an agent carry forward from past interactions, and what should it leave behind, so that it becomes more aligned with a user?
        \vspace{2px}
    \item \textbf{\textit{\underline{R}ecalibration alignment:}} How do people update their trust, expectation, and behavior as they learn what agents can(not) do?
\end{itemize}

The first three pillars, \spear{S}, \spear{P}, and \spear{E}, adopted from Terry et al.~\cite{terry2023interactive}, form the core loop of interactive alignment; the last two, \spear{A} and \spear{R}, extend this loop over time, capturing how agents adapt to people and how people recalibrate to agents~\cite{chen_align_2026,10.1145/3706599.3716291}.

We argue that these questions are not entirely new. HCI has studied closely related problems for decades under the names of grounding, activity theory, gulfs of execution and evaluation, distributed cognition, mixed-initiative interaction, adaptation, appropriation, social interaction, and groupware. What is new is the scale, opacity, agency, and adaptivity of contemporary AI systems, especially agents. This creates an opportunity to translate HCI theories and interaction mechanisms into the language of AI alignment.

\section{SPEAR: Pricinples for Interactive Human-Agent Alignment}

The five alignment questions above provide a bridge between the alignment frameworks and older foundations in HCI and related fields (as \underline{underlined}). In this section, we map each question to a corresponding HCI theoretical foundation and a set of interaction mechanisms that can support alignment in deployed agents.

\paragraph{\textbf{Specification Alignment}: Intent Grounding}
User goals are often partial, ambiguous, or context-dependent. \underline{Grounding and repair} provide a foundation here: communication requires participants to establish common ground through clarification, acknowledgment, evidence, and repair~\cite{clark1991grounding,brennan1998grounding}. Applied to agents, this suggests that alignment requires more than instruction following. Systems need mechanisms for asking clarifying questions, surfacing assumptions, confirming interpretations, and repairing misunderstandings. Grounding further distinguishes acknowledging a request from
establishing evidence of understanding~\cite{clark1991grounding}.
For example, a user at Columbia University asks for an
``\textit{affordable and convenient}'' flight to the Washington, DC,
area for HCOMP 2026 in Alexandria, Virginia. A travel-booking agent
could ask, ``\textit{Would an LGA--DCA flight under \$400, departing
between 9 a.m.\ and 6 p.m., work for you? LaGuardia is relatively
close to Columbia than JFK or EWR, while DCA is closer to
Alexandria than IAD.}'' The user can then repair this interpretation by responding,
``\textit{JFK is fine if it saves at least \$100},'' prompting the
agent to revise its proposed airport-selection criteria and
seek confirmation.
Such provisional interpretations make assumptions about price,
timing, and airport choice inspectable, allowing users to correct
the specification before the agent commits to action.

\paragraph{\textbf{Process Alignment}: Agency Coordination}
As agents act on users' behalf, alignment failures may occur even
when the final task is completed successfully. For example, a user asks, ``\textit{Book my flight for the HCOMP conference.}'' The agent finds a suitable flight and attempts to use the company payment account, but discovers that the account has expired. It then finds a connected account on the booking page: the user's saved personal card. Without asking for confirmation, the agent uses the personal card to complete the booking and reports only, ``\textit{Your flight is booked. All set.}'' The task is completed, but the user remains unaware of the payment substitution. The agent has implicitly treated authorization to book the flight as authorization to choose how to pay for it.
HCI work on \underline{distributed cognition}
and \underline{mixed-initiative interaction} provides foundations
for distributed cognitive work and the coordination
of initiative~\cite{hutchins1996distributed,hollan2000distributed,horvitz1999principles}.
AI agents allow users to offload parts of planning and decision-making,
distributing cognitive work across people, models, tools, and external
representations. However, offloading cognitive work does not by itself
determine which decisions an agent is authorized to make. Complementing
this account, the \underline{activity theory} foregrounds how tool-mediated
activity is organized through purposes, rules, community, and a
division of labor~\cite{activitytheory1, activitytheory2}.  It  adds a
question of legitimate delegation to the distribution of cognitive
work: which decisions may an agent make, under whose authority, and with
what responsibilities? Initiative is thus an alignment decision:
acting, asking, deferring, or pausing should reflect not only task
uncertainty but also authorization boundaries. Recent work on
proactive agents and ambiguity-sensitive pausing illustrates how
interaction mechanisms can support user control while enabling
automation~\cite{10.1145/3746059.3747797,zargham2022proactivity}.

\paragraph{\textbf{Evaluation Alignment}: Success Inspectability}
Users evaluate agents through the evidence available to them, yet agents
often obscure their capabilities, limitations, action
histories, and resulting effects. Norman's
\underline{gulfs of evaluation and execution} clarify why making
this evidence visible must support both judgment and
intervention~\cite{hutchins1985direct}.
For example, after booking the flight, the agent may
report, ``\textit{Your flight is confirmed.}'' This confirms task
completion but does not establish whether the booking meets the
user's constraints. More useful feedback would state,
``\textit{Your LGA--DCA flight costs \$350 and arrives at 12:30 p.m.,
30 minutes later than your requested arrival deadline.}''
Connecting the outcome to the user's budget and timing constraints
helps bridge the gulf of evaluation. Clearly labeled options such
as ``\textit{Find an earlier flight}'' and
``\textit{View cancellation terms}'' help bridge the gulf of
execution by making corrective actions discoverable.
These gulfs provide a theoretical foundation for work on
\underline{HAI interaction guidelines, explainability,
and agent visibility}, emphasizing the need for helping users understand
system capabilities, ongoing actions, and opportunities for
correction~\cite{amershi2019guidelines,yang2020reexamining,chan2024visibility}.
Visibility into agent state, action traces, goal representations,
and side effects is therefore an alignment mechanism when it
enables users to compare outcomes with their intentions and
respond to mismatches, rather than merely exposing more information.

\paragraph{\textbf{Adaptation Alignment}: Over Time}
Alignment must remain revisable as users change their goals and
incorporate agents into everyday work. HCI work on
\underline{adaptation and appropriation} shows how users repurpose
technologies and develop new practices through
use~\cite{dix2007designing,dourish2003appropriation,moran2002everyday}.
\underline{Structuration theory}
describes rules and resources as both shaping action and being
reproduced or transformed through it~\cite{stones2017structuration,giddens2014structuration}. Orlikowski's practice lens
applies this relationship to recurring technology
use~\cite{orlikowski2000structuration}.
For example, while planning travel to HCOMP 2026, a user says,
``\textit{JFK is fine if it saves at least \$100.}''
If saved and repeatedly accepted, this situational trade-off may
become a default for future trips, even when shorter ground travel
becomes more important. Similarly, when a company payment account
is unavailable, the user may authorize a temporary workaround:
``\textit{Use my personal card this time; I will request reimbursement.}''
If the agent subsequently proposes that card and the user
repeatedly accepts it for convenience, the workaround may become
routine, leaving the user responsible for advancing funds and
requesting reimbursement. In both cases, system defaults shape
subsequent actions, while repeated acceptance reinforces those
defaults without necessarily establishing their continued
appropriateness. This lens motivates editable alignment
artifacts---preferences, policies, permissions, and histories---and
opportunities to reconsider their persistence:
``\textit{Keep the \$100 airport trade-off for future trips?}''
or ``\textit{The company account is available again. Switch back?}''
Longitudinal evaluation should examine how such arrangements
become routine, how users appropriate agents into different roles,
and whether earlier choices remain open to
revision~\cite{10.1145/3643834.3661587,ross2023programmers}.

\paragraph{\textbf{Recalibration Alignment}: Human-to-Agent Alignment}
Alignment does not only shape agent behavior; interaction also changes how
people understand, use, and rely on agents. This bidirectional view treats
human adaptation not as alignment in itself, but as a process through
which people develop more appropriate ways of interpreting and
collaborating with agents. Works on \underline{computers as social actors}, \underline{groupware and social interaction} highlights how people
respond socially to computers and adapt their coordination practices to
technology-mediated work~\cite{nass1994computers,grudin1994groupware}.
Dennett's \underline{intentional stance} describes predicting a
system's behavior by attributing beliefs and desires to it, without
requiring that it possess human-like mental states~\cite{dennett1989intentional}.
For example, after several successful travel bookings, a user may
reason, ``\textit{It knows my preferences and wants to find what works
best for me.}'' Such interactions can reshape:
\begin{itemize}[leftmargin=10pt]
    \item expectations (``\textit{It should check baggage fees without being asked}''),
\item trust (``\textit{Its previous recommendations were good, so I trust this}''),
    \item mental models (``\textit{It will now remember my \$100 airport trade-off}''),
\item practices (``\textit{I now check only the summary, not the full itinerary}''),
    \item delegation boundaries (``\textit{Book flights under \$400 without asking me each time}'').
\end{itemize}

\noindent These adaptations do not necessarily constitute improved
alignment. Alignment concerns whether users' evolving expectations,
mental models, and reliance become \textit{calibrated to the system's
actual capabilities, behavior, and limitations}. A user who comes to
expect persistent memory when the system cannot reliably retain
preferences has adapted to the system, but in a way that creates a
mismatch between their mental model and the agent. Capability
demonstrations and expectation-setting mechanisms~\cite{10.1145/3290605.3300641},
including explicit memory boundaries such as ``\textit{Your airport
preference applies to this booking only; it has not been saved for
future trips},'' can help users calibrate their interpretations and
reliance. Human-to-agent alignment should therefore support
\textit{informed and revisable calibration}, rather than assume that
greater trust, less checking, or broader delegation is necessarily
beneficial.



\section{Summary: Interaction Mechanisms as Alignment Mechanisms}

Across these five questions, we argue that interaction design mechanisms can be understood as alignment mechanisms. Clarification, pausing, visibility, onboarding, correction, explanation, customization, and expectation-setting are not merely interface details; they are the means by which alignment is negotiated, maintained, and repaired after deployment.


We see our work contributes a bridge between AI alignment, human computation, and human-AI interaction. Human feedback is central to modern alignment pipelines, but the role of humans is often compressed into labels, rankings, demonstrations, or evaluations. An HCI perspective expands the role of humans from feedback providers to situated collaborators, evaluators, appropriators, and co-adapting participants in sociotechnical systems. This shift is especially important for agentic systems that act over time, coordinate across people and tools, and shape collective work.


\bibliographystyle{ACM-Reference-Format}
\bibliography{references}

@incollection{giddens2014structuration,
  title={Structuration theory: past, present and future},
  author={Giddens, Anthony},
  booktitle={Giddens' theory of structuration},
  pages={201--221},
  year={2014},
  publisher={Routledge}
}

@article{orlikowski2000structuration,
  title={Using technology and constituting structures: A practice lens for studying technology in organizations},
  author={Orlikowski, Wanda J},
  journal={Organization science},
  volume={11},
  number={4},
  pages={404--428},
  year={2000},
  publisher={Informs}
}

@inproceedings{10.1145/3290605.3300641,
author = {Kocielnik, Rafal and Amershi, Saleema and Bennett, Paul N.},
title = {Will You Accept an Imperfect AI? Exploring Designs for Adjusting End-user Expectations of AI Systems},
year = {2019},
isbn = {9781450359702},
publisher = {Association for Computing Machinery},
address = {New York, NY, USA},
url = {https://doi.org/10.1145/3290605.3300641},
doi = {10.1145/3290605.3300641},
booktitle = {Proceedings of the 2019 CHI Conference on Human Factors in Computing Systems},
pages = {1–14},
numpages = {14},
location = {Glasgow, Scotland Uk},
series = {CHI '19}
}

@book{dennett1989intentional,
  title={The intentional stance},
  author={Dennett, Daniel C},
  year={1989},
  publisher={MIT press}
}

@book{stones2017structuration,
  title={Structuration theory},
  author={Stones, Rob},
  year={2017},
  publisher={Bloomsbury Publishing}
}

@article{hutchins1985direct,
  title={Direct manipulation interfaces},
  author={Hutchins, Edwin L and Hollan, James D and Norman, Donald A},
  journal={Human--computer interaction},
  volume={1},
  number={4},
  pages={311--338},
  year={1985},
  publisher={Taylor \& Francis}
}

@article{activitytheory1,
  title={Activity theory and human-computer interaction},
  author={Nardi, Bonnie A},
  journal={Context and consciousness: Activity theory and human-computer interaction},
  volume={436},
  number={1},
  pages={7--16},
  year={1996}
}

@article{activitytheory2,
  title={Activity theory as a framework for analyzing and redesigning work},
  author={Engestrom, Yrjo},
  journal={Ergonomics},
  volume={43},
  number={7},
  pages={960--974},
  year={2000},
  publisher={Taylor \& Francis}
}

@inproceedings{christiano2017deep,
author = {Christiano, Paul F. and Leike, Jan and Brown, Tom B. and Martic, Miljan and Legg, Shane and Amodei, Dario},
title = {Deep reinforcement learning from human preferences},
year = {2017},
isbn = {9781510860964},
publisher = {Curran Associates Inc.},
address = {Red Hook, NY, USA},
booktitle = {Proceedings of the 31st International Conference on Neural Information Processing Systems},
pages = {4302–4310},
numpages = {9},
location = {Long Beach, California, USA},
series = {NIPS'17}
}

@inproceedings{ouyang2022training,
author = {Ouyang, Long and Wu, Jeff and Jiang, Xu and Almeida, Diogo and Wainwright, Carroll L. and Mishkin, Pamela and Zhang, Chong and Agarwal, Sandhini and Slama, Katarina and Ray, Alex and Schulman, John and Hilton, Jacob and Kelton, Fraser and Miller, Luke and Simens, Maddie and Askell, Amanda and Welinder, Peter and Christiano, Paul and Leike, Jan and Lowe, Ryan},
title = {Training language models to follow instructions with human feedback},
year = {2022},
isbn = {9781713871088},
publisher = {Curran Associates Inc.},
address = {Red Hook, NY, USA},
booktitle = {Proceedings of the 36th International Conference on Neural Information Processing Systems},
articleno = {2011},
numpages = {15},
location = {New Orleans, LA, USA},
series = {NIPS '22}
}

@misc{bai2022constitutional,
      title={Constitutional AI: Harmlessness from AI Feedback}, 
      author={Yuntao Bai and Saurav Kadavath and Sandipan Kundu and Amanda Askell and Jackson Kernion and Andy Jones and Anna Chen and Anna Goldie and Azalia Mirhoseini and Cameron McKinnon and Carol Chen and Catherine Olsson and Christopher Olah and Danny Hernandez and Dawn Drain and Deep Ganguli and Dustin Li and Eli Tran-Johnson and Ethan Perez and Jamie Kerr and Jared Mueller and Jeffrey Ladish and Joshua Landau and Kamal Ndousse and Kamile Lukosuite and Liane Lovitt and Michael Sellitto and Nelson Elhage and Nicholas Schiefer and Noemi Mercado and Nova DasSarma and Robert Lasenby and Robin Larson and Sam Ringer and Scott Johnston and Shauna Kravec and Sheer El Showk and Stanislav Fort and Tamera Lanham and Timothy Telleen-Lawton and Tom Conerly and Tom Henighan and Tristan Hume and Samuel R. Bowman and Zac Hatfield-Dodds and Ben Mann and Dario Amodei and Nicholas Joseph and Sam McCandlish and Tom Brown and Jared Kaplan},
      year={2022},
      eprint={2212.08073},
      archivePrefix={arXiv},
      primaryClass={cs.CL},
      url={https://arxiv.org/abs/2212.08073}, 
}

@misc{terry2023interactive,
  author        = {Terry, Michael and Kulkarni, Chinmay and Wattenberg, Martin and Dixon, Lucas and Morris, Meredith Ringel},
  title         = {Interactive AI Alignment: Specification, Process, and Evaluation Alignment},
  year          = {2023},
  eprint        = {2311.00710},
  archivePrefix = {arXiv}
}

@incollection{clark1991grounding,
  author    = {Clark, Herbert H. and Brennan, Susan E.},
  title     = {Grounding in Communication},
  booktitle = {Perspectives on Socially Shared Cognition},
  editor    = {Resnick, Lauren B. and Levine, John M. and Teasley, Stephanie D.},
  pages     = {127--149},
  publisher = {American Psychological Association},
  address   = {Washington, DC},
  year      = {1991},
  doi       = {10.1037/10096-006}
}

@incollection{brennan1998grounding,
  author    = {Brennan, Susan E.},
  title     = {The Grounding Problem in Conversations with and through Computers},
  booktitle = {Social and Cognitive Approaches to Interpersonal Communication},
  editor    = {Fussell, Susan R. and Kreuz, Roger J.},
  pages     = {201--225},
  publisher = {Lawrence Erlbaum Associates},
  year      = {1998}
}

@incollection{hutchins1996distributed,
  author    = {Hutchins, Edwin and Klausen, Tove},
  title     = {Distributed Cognition in an Airline Cockpit},
  booktitle = {Cognition and Communication at Work},
  editor    = {Engestr{\"o}m, Yrj{\"o} and Middleton, David},
  pages     = {15--34},
  publisher = {Cambridge University Press},
  address   = {Cambridge},
  year      = {1996}
}

@article{hollan2000distributed,
author = {Hollan, James and Hutchins, Edwin and Kirsh, David},
title = {Distributed cognition: toward a new foundation for human-computer interaction research},
year = {2000},
issue_date = {June 2000},
publisher = {Association for Computing Machinery},
address = {New York, NY, USA},
volume = {7},
number = {2},
issn = {1073-0516},
url = {https://doi.org/10.1145/353485.353487},
doi = {10.1145/353485.353487},
journal = {ACM Trans. Comput.-Hum. Interact.},
month = jun,
pages = {174–196},
numpages = {23}
}

@inproceedings{horvitz1999principles,
author = {Horvitz, Eric},
title = {Principles of mixed-initiative user interfaces},
year = {1999},
isbn = {0201485591},
publisher = {Association for Computing Machinery},
address = {New York, NY, USA},
url = {https://doi.org/10.1145/302979.303030},
doi = {10.1145/302979.303030},
booktitle = {Proceedings of the SIGCHI Conference on Human Factors in Computing Systems},
pages = {159–166},
numpages = {8},
location = {Pittsburgh, Pennsylvania, USA},
series = {CHI '99}
}

@inproceedings{amershi2019guidelines,
author = {Amershi, Saleema and Weld, Dan and Vorvoreanu, Mihaela and Fourney, Adam and Nushi, Besmira and Collisson, Penny and Suh, Jina and Iqbal, Shamsi and Bennett, Paul N. and Inkpen, Kori and Teevan, Jaime and Kikin-Gil, Ruth and Horvitz, Eric},
title = {Guidelines for Human-AI Interaction},
year = {2019},
isbn = {9781450359702},
publisher = {Association for Computing Machinery},
address = {New York, NY, USA},
url = {https://doi.org/10.1145/3290605.3300233},
doi = {10.1145/3290605.3300233},
booktitle = {Proceedings of the 2019 CHI Conference on Human Factors in Computing Systems},
pages = {1–13},
numpages = {13},
location = {Glasgow, Scotland Uk},
series = {CHI '19}
}

@inproceedings{moran2002everyday,
author = {Moran, Thomas P},
title = {Everyday adaptive design},
year = {2002},
isbn = {1581135157},
publisher = {Association for Computing Machinery},
address = {New York, NY, USA},
url = {https://doi.org/10.1145/778712.778715},
doi = {10.1145/778712.778715},
booktitle = {Proceedings of the 4th Conference on Designing Interactive Systems: Processes, Practices, Methods, and Techniques},
pages = {13–14},
numpages = {2},
location = {London, England},
series = {DIS '02}
}

@article{dourish2003appropriation,
author = {Dourish, Paul},
title = {The Appropriation of Interactive Technologies: Some Lessons from Placeless Documents},
year = {2003},
issue_date = {2003},
publisher = {Kluwer Academic Publishers},
address = {USA},
volume = {12},
number = {4},
issn = {0925-9724},
url = {https://doi.org/10.1023/A:1026149119426},
doi = {10.1023/A:1026149119426},
journal = {Comput. Supported Coop. Work},
month = sep,
pages = {465–490},
numpages = {26}
}

@inproceedings{dix2007designing,
author = {Dix, Alan},
title = {Designing for appropriation},
year = {2007},
isbn = {9781902505954},
publisher = {BCS Learning \& Development Ltd.},
address = {Swindon, GBR},
booktitle = {Proceedings of the 21st British HCI Group Annual Conference on People and Computers: HCI...but Not as We Know It - Volume 2},
pages = {27–30},
numpages = {4},
location = {University of Lancaster, United Kingdom},
series = {BCS-HCI '07}
}

@inproceedings{nass1994computers,
author = {Nass, Clifford and Steuer, Jonathan and Tauber, Ellen R.},
title = {Computers are social actors},
year = {1994},
isbn = {0897916506},
publisher = {Association for Computing Machinery},
address = {New York, NY, USA},
url = {https://doi.org/10.1145/191666.191703},
doi = {10.1145/191666.191703},
booktitle = {Proceedings of the SIGCHI Conference on Human Factors in Computing Systems},
pages = {72–78},
numpages = {7},
location = {Boston, Massachusetts, USA},
series = {CHI '94}
}

@article{grudin1994groupware,
author = {Grudin, Jonathan},
title = {Groupware and social dynamics: eight challenges for developers},
year = {1994},
issue_date = {Jan. 1994},
publisher = {Association for Computing Machinery},
address = {New York, NY, USA},
volume = {37},
number = {1},
issn = {0001-0782},
url = {https://doi.org/10.1145/175222.175230},
doi = {10.1145/175222.175230},
journal = {Commun. ACM},
month = jan,
pages = {92–105},
numpages = {14}
}

@inproceedings{10.1145/3706599.3716291,
author = {Shen, Hua and Knearem, Tiffany and Ghosh, Reshmi and Liu, Michael Xieyang and Monroy-Hern\'{a}ndez, Andr\'{e}s and Wu, Tongshuang and Yang, Diyi and Huang, Yun and Mitra, Tanushree and Li, Yang and Hearst, Marti},
title = {Bidirectional Human-AI Alignment: Emerging Challenges and Opportunities},
year = {2025},
isbn = {9798400713958},
publisher = {Association for Computing Machinery},
address = {New York, NY, USA},
url = {https://doi.org/10.1145/3706599.3716291},
doi = {10.1145/3706599.3716291},
booktitle = {Proceedings of the Extended Abstracts of the CHI Conference on Human Factors in Computing Systems},
articleno = {857},
numpages = {6},
location = {
},
series = {CHI EA '25}
}

@inproceedings{10.1145/3643834.3661587,
author = {Long, Tao and Gero, Katy Ilonka and Chilton, Lydia B},
title = {Not Just Novelty: A Longitudinal Study on Utility and Customization of an AI Workflow},
year = {2024},
isbn = {9798400705830},
publisher = {Association for Computing Machinery},
address = {New York, NY, USA},
url = {https://doi.org/10.1145/3643834.3661587},
doi = {10.1145/3643834.3661587},
booktitle = {Proceedings of the 2024 ACM Designing Interactive Systems Conference},
pages = {782–803},
numpages = {22},
location = {Copenhagen, Denmark},
series = {DIS '24}
}

@inproceedings{10.1145/3746059.3747797,
author = {Peng, Yi-Hao and Li, Dingzeyu and Bigham, Jeffrey P and Pavel, Amy},
title = {Morae: Proactively Pausing UI Agents for User Choices},
year = {2025},
isbn = {9798400720376},
publisher = {Association for Computing Machinery},
address = {New York, NY, USA},
url = {https://doi.org/10.1145/3746059.3747797},
doi = {10.1145/3746059.3747797},
booktitle = {Proceedings of the 38th Annual ACM Symposium on User Interface Software and Technology},
articleno = {198},
numpages = {14},
location = {
},
series = {UIST '25}
}

@inproceedings{zargham2022proactivity,
 author = {Zargham, Nima and Reicherts, Leon and Bonfert, Michael and Voelkel, Sarah Theres and Schoening, Johannes and Malaka, Rainer and Rogers, Yvonne},
title = {Understanding Circumstances for Desirable Proactive Behaviour of Voice Assistants: The Proactivity Dilemma},
year = {2022},
isbn = {9781450397391},
publisher = {Association for Computing Machinery},
address = {New York, NY, USA},
url = {https://doi.org/10.1145/3543829.3543834},
doi = {10.1145/3543829.3543834},
booktitle = {Proceedings of the 4th Conference on Conversational User Interfaces},
articleno = {3},
numpages = {14},
location = {Glasgow, United Kingdom},
series = {CUI '22}
}

@inproceedings{yang2020reexamining,
author = {Yang, Qian and Steinfeld, Aaron and Ros\'{e}, Carolyn and Zimmerman, John},
title = {Re-examining Whether, Why, and How Human-AI Interaction Is Uniquely Difficult to Design},
year = {2020},
isbn = {9781450367080},
publisher = {Association for Computing Machinery},
address = {New York, NY, USA},
url = {https://doi.org/10.1145/3313831.3376301},
doi = {10.1145/3313831.3376301},
booktitle = {Proceedings of the 2020 CHI Conference on Human Factors in Computing Systems},
pages = {1–13},
numpages = {13},
location = {Honolulu, HI, USA},
series = {CHI '20}
}

@inproceedings{chan2024visibility,
author = {Chan, Alan and Ezell, Carson and Kaufmann, Max and Wei, Kevin and Hammond, Lewis and Bradley, Herbie and Bluemke, Emma and Rajkumar, Nitarshan and Krueger, David and Kolt, Noam and Heim, Lennart and Anderljung, Markus},
title = {Visibility into AI Agents},
year = {2024},
isbn = {9798400704505},
publisher = {Association for Computing Machinery},
address = {New York, NY, USA},
url = {https://doi.org/10.1145/3630106.3658948},
doi = {10.1145/3630106.3658948},
booktitle = {Proceedings of the 2024 ACM Conference on Fairness, Accountability, and Transparency},
pages = {958–973},
numpages = {16},
location = {Rio de Janeiro, Brazil},
series = {FAccT '24}
}

@misc{chen_align_2026,
	title = {Align {AI} to {Dynamic} {Human}-{AI} {Workflows}},
	url = {http://arxiv.org/abs/2607.14240},
	doi = {10.48550/arXiv.2607.14240},
	urldate = {2026-09-17},
	publisher = {arXiv},
	author = {Chen, Valerie and Gonzalez, Cleotilde and Woolley, Anita Williams and Lee, Michael and Wu, Tongshuang and Conitzer, Vincent and Singh, Aarti},
	month = jul,
	year = {2026},
	note = {arXiv:2607.14240 [cs.AI]
version: 1},
}

@inproceedings{ross2023programmers,
author = {Ross, Steven I. and Martinez, Fernando and Houde, Stephanie and Muller, Michael and Weisz, Justin D.},
title = {The Programmer’s Assistant: Conversational Interaction with a Large Language Model for Software Development},
year = {2023},
isbn = {9798400701061},
publisher = {Association for Computing Machinery},
address = {New York, NY, USA},
url = {https://doi.org/10.1145/3581641.3584037},
doi = {10.1145/3581641.3584037},
booktitle = {Proceedings of the 28th International Conference on Intelligent User Interfaces},
pages = {491–514},
numpages = {24},
location = {Sydney, NSW, Australia},
series = {IUI '23}
}






\end{document}